\documentclass{llncs}
\usepackage{amsmath} 
\usepackage{hyperref}
\usepackage{version}
\usepackage{pcldb}

\begin{document}

\title{Paraconsistent Logic and \\
       Query Answering in Inconsistent Databases}
\author{C.~A. Middelburg \\
        {\small ORCID: \url{https://orcid.org/0000-0002-8725-0197}}}
\institute{Informatics Institute, Faculty of Science, University of
           Amsterdam, \\
           Science Park~900, 1098~XH Amsterdam, the Netherlands \\
           \email{C.A.Middelburg@uva.nl}}

\maketitle

\begin{abstract}
This paper concerns the paraconsistent logic \foLPif\ and an application
of it in the area of relational database theory.
The notions of a relational database, a query applicable to a relational 
database, and a consistent answer to a query with respect to a possibly 
inconsistent relational database are considered from the perspective of 
this logic.
This perspective enables among other things the definition of a 
consistent answer to a query with respect to a possibly inconsistent 
database without resort to database repairs.
In a previous paper, \foLPif\ is presented with a sequent-style natural 
deduction proof system.
In this paper, a sequent calculus proof system is presented because it 
is common to use a sequent calculus proof system as the basis of proof 
search procedures and such procedures may form the core of algorithms for 
computing consistent answers to queries.

\begin{keywords} 
relational database; inconsistent database; consistent query answering; 
paraconsistent logic; sequent calculus
\end{keywords}

\end{abstract}

\section{Introduction}
\label{sect-intro}

In the area of relational database theory, rather often the view is 
taken in which a database is a theory of first-order classical logic, a 
query is a formula, and query answering amounts to proving in 
first-order classical logic that a formula is a logical consequence of 
a theory.
In~\cite{Rei84a}, the term \emph{proof-theoretic view} is introduced 
for this view and various arguments in favor of this view are given. 
In work on query answering in inconsistent databases based on this view, 
resort to (consistent) repairs of inconsistent databases is considered
unavoidable to come to a notion of a consistent answer to a possibly 
inconsistent database (see e.g.~\cite{ABC99a}).
The reason for this is that in classical logic every formula is a 
logical consequence of an inconsistent theory.

In~\cite{Bry97a}, the resort to repairs is avoided by switching from
first-order classical logic to first-order minimal logic, a logic in 
which not every formula is a logical consequence of an inconsistent 
theory.
By some shortcomings in~\cite{Bry97a}, there has been no follow-up of 
this work.
The main shortcoming is that a semantics with respect to which the 
presented proof system is sound and complete is not given.
By that, it remains unclear how the work fits the existing (concrete or 
abstract) views on what is a database.
Actually, there exists a Kripke semantics of the proposional fragment
(see e.g.~\cite{CJV17a}), but that semantics seems difficult to relate 
the existing views on what is a database.

This paper considers consistent query answering from the perspective of
\foLPif, another first-order logic in which not every formula is a 
logical consequence of an inconsistent theory.
A sequent calculus proof system of \foLPif\ and a three-valued semantics 
with respect to which the proof system is sound and complete are given.
The notions of a relational database, a query applicable to a relational
database, and an answer to a query with respect to a relational database 
are defined in the setting of \foLPif.
The definitions concerned are based on those given in~\cite{Rei84a}.
Two notions of a consistent answer to a query with respect to a possibly
inconsistent relational database are introduced.
One of them is reminiscent of the notion of a consistent answer 
from~\cite{Bry97a} and the other is essentially the same as the notion 
of a consistent answer from~\cite{ABC99a}.

Proof search procedures may form the core of algorithms for computing 
consistent answers to queries.
It is common to use a sequent calculus proof system as the basis of 
proof search procedures.
That is why a sequent calculus proof system of \foLPif\ is presented in 
this paper.
The proof system of first-order minimal logic presented in~\cite{Bry97a} 
is a natural deduction proof system.
A natural deduction proof system can also be used as the basis of a 
proof search procedure, but it is not so widely known how this can be 
done.  
The lack of any remark about a proof search procedure for first-order 
minimal logic is sometimes considered a shortcoming in~\cite{Bry97a} 
as well.

A logic is called a paraconsistent logic if in the logic not every 
formula is a logical consequence of an inconsistent theory.
In~\cite{Pri79a}, Priest proposed the paraconsistent propositional logic
LP (Logic of Paradox) and its first-order extension LPQ.\,
\foLPif\ is LPQ enriched with a falsity constant and an implication 
connective for which the standard deduction theorem holds. 
\foLPif\ is essentially the same as \foJiii~\cite{DOt85a} and 
\foLPc~\cite{Pic18a}.
In~\cite{Mid20c}, a sequent-style natural deduction proof system for 
\foLPif\ is presented.
Several main properties of the logical consequence relation and the 
logical equivalence relation of \foLPif\ are also treated in that paper.
 
In \foLPif, for every inconsistent theory $\Gamma$ in which the falsity 
constant $\False$ does not occur, for every formula $A$ that does not 
have function symbols, predicate symbols or free variables in common 
with $\Gamma$, $A$ is a logical consequence of $\Gamma$ only if $A$ is 
a logical consequence of the empty theory.
In minimal logic, for every inconsistent theory $\Gamma$, for every 
formula $A$, $\Not A$ is a logical consequence of~$\Gamma$.
Therefore, \foLPif\ is considered a genuine paraconsistent logic and 
minimal logic is not considered a genuine paraconsistent logic
(cf.~\cite{Odi07a}).
Moreover, the properties of \foLPif\ treated in~\cite{Mid20c} indicate 
among other things that the logical consequence relation and the logical 
equivalence relation of \foLPif\ are very close to those of classical 
logic.
That is why the choice has been made to consider in this paper query 
answering in inconsistent databases from the perspective of \foLPif.

The structure of this paper is as follows.
First, the language of \foLPif, a sequent calculus proof system of 
\foLPif, and a three-valued semantics of \foLPif\ are presented 
(Sections~\ref{LANGUAGE}, \ref{PROOF-RULES}, and~\ref{INTERPRETATION}).
Next, relational databases and query answering in possibly inconsistent
relational databases are considered from the perspective of \foLPif\
\linebreak[2] (Sections~\ref{DATABASE} and~\ref{QUERY-ANSWERING}).
After that, examples of query answering are given 
(Section~\ref{EXAMPLES}) and some remaining remarks about consistent 
query answering are made (Section~\ref{REMARKS}).
Finally, some concluding remarks are made (Section~\ref{CONCLUSIONS}).

In order to make this paper self-contained, large parts of 
Sections~\ref{LANGUAGE} and~\ref{INTERPRETATION} have been copied near 
verbatim or slightly modified from~\cite{Mid20c}.

\section{The Language of \foLPif}
\label{LANGUAGE}

In this section the language of the paraconsistent logic \foLPif\ is 
described.
First the notion of a signature is introduced and then the terms and 
formulas of \foLPif\ are defined for a fixed but arbitrary signature. 
Moreover, some relevant notational conventions and abbreviations are 
presented and some remarks about free variables and substitution are 
made.
In coming sections, the proof system of \foLPif\ and the interpretation 
of the terms and formulas of \foLPif\ are defined for a fixed but 
arbitrary signature.

\subsubsection*{Signatures}

It is assumed that the following has been given:
(a)~a countably infinite set $\SVar$ of \emph{variables},
(b)~for each $n \in \Nat$, a countably infinite set $\SFunc{n}$ of 
\emph{function symbols of arity $n$}, and,
(c)~for each $n \in \Nat$, a countably infinite set $\SPred{n}$ of 
\emph{predicate symbols of arity $n$}.
It is also assumed that all these sets and the set $\set{=}$ are 
mutually disjoint.
We write $\SSymb$ for the set
$\SVar \union \Union \set{\SFunc{n} \where n \in \Nat} \union
 \Union \set{\SPred{n} \where n \in \Nat}$.

Function symbols of arity $0$ are also known as \emph{constant symbols} 
and predicate symbols of arity $0$ are also known as \emph{proposition 
symbols}.

A \emph{signature} $\Sigma$ is a subset of $\SSymb \diff \SVar$.
We write $\Func{n}(\Sigma)$ and $\Pred{n}(\Sigma)$, 
where $\Sigma$ is a signature and $n \in \Nat$, 
for the sets $\Sigma \inter \SFunc{n}$ and $\Sigma \inter \SPred{n}$, 
respectively.

The language of \foLPif\ will be defined for a fixed but arbitrary 
signature~$\Sigma$.
This language will be called the language of \foLPif\ over $\Sigma$ or 
shortly the language of $\foLPif(\Sigma)$.
The corresponding proof system and interpretation will be called the
proof system of $\foLPif(\Sigma)$ and the interpretation of 
$\foLPif(\Sigma)$.

\subsubsection*{Terms and formulas}

The language of $\foLPif(\Sigma)$ consists of terms and formulas.
They are constructed according to the formation rules given below.

The set of all \emph{terms of $\foLPif(\Sigma)$}, 
written $\STerm{\Sigma}$, is inductively defined by the following 
formation rules:
\begin{enumerate}
\item
if $x \in \SVar$, then $x \in \STerm{\Sigma}$;
\item
if $c \in \Func{0}(\Sigma)$, then $c \in \STerm{\Sigma}$;
\item
if $f \in \Func{n+1}(\Sigma)$ and 
$t_1,\ldots,t_{n+1} \in \STerm{\Sigma}$, then 
$f(t\sb1,\ldots,t_{n+1}) \in \STerm{\Sigma}$.
\end{enumerate}
The set of all \emph{closed terms of $\foLPif(\Sigma)$} is the subset 
of $\SForm{\Sigma}$ inductively defined by the formation rules 2 and 3.

The set of all \emph{formulas of $\foLPif(\Sigma)$}, 
written $\SForm{\Sigma}$, is inductively defined by the following 
formation rules:
\begin{enumerate}
\item
$\False \in \SForm{\Sigma}$;
\item
if $p \in \Pred{0}(\Sigma)$, then 
$p \in \SForm{\Sigma}$;
\item
if $P \in \Pred{n+1}(\Sigma)$ and 
$t_1,\ldots,t_{n+1} \in \STerm{\Sigma}$, then 
$P(t_1,\ldots,t_{n+1}) \in \SForm{\Sigma}$;
\item
if $t_1,t_2 \in \STerm{\Sigma}$, then $t_1 = t_2 \in \SForm{\Sigma}$;
\item
if $A \in \SForm{\Sigma}$, then $\Not A \in \SForm{\Sigma}$;
\item
if $A_1,A_2 \in \SForm{\Sigma}$, then 
$A_1 \CAnd A_2,\, A_1 \COr A_2,\, A_1 \IImpl A_2 \in \SForm{\Sigma}$;
\item
if $x \in \SVar$ and $A \in \SForm{\Sigma}$, then 
$\CForall{x}{A},\, \CExists{x}{A} \in \SForm{\Sigma}$.
\end{enumerate}
The set of all \emph{atomic formulas of $\foLPif(\Sigma)$} is the subset 
of $\SForm{\Sigma}$ inductively defined by the formation rules 1--4.
The set of all \emph{literals of $\foLPif(\Sigma)$} is the subset of 
$\SForm{\Sigma}$ inductively defined by the formation rules 1--5.

For the connectives $\Not$, $\CAnd$, $\COr$, and $\IImpl$ and the 
quantifiers $\forall$ and $\exists$, the classical truth-conditions and 
falsehood-conditions are retained.
Except for implications, a formula is classified as both-true-and-false 
exactly when it cannot be classified as true or false by these 
conditions.

We write $e_1 \equiv e_2$, where $e_1$ and $e_2$ are terms from 
$\STerm{\Sigma}$ or formulas from $\SForm{\Sigma}$, to indicate that
$e_1$ is syntactically equal to $e_2$.

\subsubsection*{Notational conventions and abbreviations}

The following will sometimes be used without mentioning (with or without 
decoration):
$x$~as a meta-variable ranging over all variables from 
$\SVar$,
$t$~as a meta-variable ranging over all terms from 
$\STerm{\Sigma}$, 
$A$ as a meta-variable ranging over all formulas from 
$\SForm{\Sigma}$, and
$\Gamma$ as a meta-variable ranging over all finite sets of 
formulas from~$\SForm{\Sigma}$.

The string representation of terms and formulas suggested by the 
formation rules given above can lead to syntactic ambiguities. 
Parentheses are used to avoid  such ambiguities.
The need to use parentheses is reduced by ranking the precedence of the 
logical connectives $\Not$, $\CAnd$, $\COr$, $\IImpl$.
The enumeration presents this order from the highest precedence to the
lowest precedence.
Moreover, the scope of the quantifiers extends as far as possible to
the right and $\CForall{x_1}{\cdots \CForall{x_n}{A}}$ is usually written 
as $\CForall{x_1,\ldots,x_n}{A}$.

The following abbreviation is used:
$\True$ stands for $\Not \False$.

\subsubsection*{Free variables and substitution}

Free variables of a term or formula and substitution for variables in a 
term or formula are defined in the usual way.

Let $x$ be a variable from $\SVar$, $t$ be a term from 
$\STerm{\Sigma}$, and $e$ be a term from $\STerm{\Sigma}$ or a formula 
from $\SForm{\Sigma}$.
Then we write $\subst{x \assign t} e$ for the result of substituting the 
term $t$ for the free occurrences of the variable $x$ in~$e$, 
avoiding (by means of renaming of bound variables) free variables 
becoming bound in $t$.

\section{A Proof System of $\foLPif(\Sigma)$}
\label{PROOF-RULES}

In this section, a sequent calculus proof system of $\foLPif(\Sigma)$ is 
presented.
This means that the inference rules have sequents as premises and 
conclusions.
First, the notion of a sequent is introduced. 
Then, the inference rules of the proof system of $\foLPif(\Sigma)$ are 
presented.
After that, the notion of a derivation of a sequent from a set of 
sequents and the notion of a proof of a sequent are introduced.
An extension of the proof system of $\foLPif(\Sigma)$ which can serve as 
a proof system for first-order classical logic is also described.

\subsubsection*{Sequents}

In $\foLPif(\Sigma)$, a \emph{sequent} is an expression of the form 
$\Gamma \scEnt \Delta$, where $\Gamma$ and $\Delta$ are finite sets of 
formulas from $\SForm{\Sigma}$.
We write $\Gamma,\Gamma'$ for $\Gamma \union \Gamma'$ and 
$A$, where $A$ is a formula from $\SForm{\Sigma}$, for $\set{A}$ on both 
sides of a sequent.
Moreover, we write ${} \scEnt \Delta$ instead of 
$\emptyset \scEnt \Delta$. 

A sequent $\Gamma \scEnt \Delta$ states that the logical consequence 
relation that is defined in Section~\ref{INTERPRETATION} holds between 
$\Gamma$ and $\Delta$.
Informally speaking, that logical consequence relation holds between 
$\Gamma$ and $\Delta$ if, whenever every formula from $\Gamma$ is not 
false, at least one formula from $\Delta$ is not false.
If a sequent $\Gamma \scEnt \Delta$ can be proved by means of the rules 
of inference given below, then that logical consequence relation holds 
between $\Gamma$ and~$\Delta$.
\pagebreak[2]

\subsubsection*{Rules of inference}

The sequent calculus proof system of $\foLPif(\Sigma)$ consists of the 
inference rules given in Table~\ref{table-proof-system}.
\begin{table}[p!]
\caption{Sequent calculus proof system of $\foLPif(\Sigma)$}
\label{table-proof-system}
\vspace*{-3ex} \par \mbox{} 
\renewcommand{\arraystretch}{1.2} 
\centering
\begin{tabular}[t]{@{}c@{}}
\hline
\begin{small}
\begin{tabular}{@{}l@{}}
\InfRuleC{I}
 {{}}
 {A, \Gamma \scEnt \Delta, A}
 {$\ast$}
\\[3ex] 
\InfRule{$\False$-L}
 {{}}
 {\False, \Gamma \scEnt \Delta}
\\[3ex] 
\InfRule{$\CAnd$-L}
 {A\sb1, A\sb2, \Gamma \scEnt \Delta}
 {A\sb1 \CAnd A\sb2, \Gamma \scEnt \Delta}
\\[3ex]
\InfRule{$\COr$-L}
 {A\sb1, \Gamma \scEnt \Delta \quad
  A\sb2, \Gamma \scEnt \Delta}
 {A\sb1 \COr A\sb2, \Gamma \scEnt \Delta}
\\[3ex]
\InfRule{$\IImpl$-L}
 {\Gamma \scEnt \Delta, A\sb1 \quad
  A\sb2, \Gamma \scEnt \Delta}
 {A\sb1 \IImpl A\sb2, \Gamma \scEnt \Delta}
\\[3ex]
\InfRule{$\forall$-L}
 {\subst{x \assign t}A, \Gamma \scEnt \Delta}
 {\CForall{x}{A}, \Gamma \scEnt \Delta}
\\[3ex]
\InfRuleC{$\exists$-L}
 {\subst{x \assign y}A, \Gamma \scEnt \Delta}
 {\CExists{x}{A}, \Gamma \scEnt \Delta}
 {\ddag}
\\[3ex]
\phantom{
\InfRule{$\Not \False$-R}
 {{}}
 {\Gamma \scEnt \Delta, \Not \False}
}
\\[3ex] 
\InfRule{$\Not \Not$-L}
 {A, \Gamma \scEnt \Delta}
 {\Not \Not A, \Gamma \scEnt \Delta}
\\[3ex]
\InfRule{$\Not \CAnd$-L}
 {\Not A\sb1, \Gamma \scEnt \Delta \quad
  \Not A\sb2, \Gamma \scEnt \Delta}
 {\Not (A\sb1 \CAnd A\sb2), \Gamma \scEnt \Delta}
\\[3ex]
\InfRule{$\Not \COr$-L}
 {\Not A\sb1, \Not A\sb2, \Gamma \scEnt \Delta}
 {\Not (A\sb1 \COr A\sb2), \Gamma \scEnt \Delta}
\\[3ex]
\InfRule{$\Not \IImpl$-L}
 {A\sb1, \Not A\sb2, \Gamma \scEnt \Delta}
 {\Not (A\sb1 \IImpl A\sb2), \Gamma \scEnt \Delta}
\\[3ex]
\InfRuleC{$\Not \forall$-L}
 {\Not \subst{x \assign y}A, \Gamma \scEnt \Delta}
 {\Not \CForall{x}{A}, \Gamma \scEnt \Delta}
 {\ddag}
\\[3ex]
\InfRule{$\Not \exists$-L}
 {\Not \subst{x \assign t}A, \Gamma \scEnt \Delta}
 {\Not \CExists{x}{A}, \Gamma \scEnt \Delta}
\\[3ex]
\InfRule{$=$-Refl}
 {t = t, \Gamma \scEnt \Delta}
 {\Gamma \scEnt \Delta}
\\[3ex]
\end{tabular}
\qquad \qquad
\begin{tabular}{@{}l@{}} 
\phantom{
\InfRule{I}
 {{}}
 {A, \Gamma \scEnt \Delta, A}
}
\\[3ex] 
\InfRuleC{$\Not$-R}
 {A, \Gamma \scEnt \Delta}
 {\Gamma \scEnt \Delta, \Not A}
{\dag}
\\[3ex]
\InfRule{$\CAnd$-R}
 {\Gamma \scEnt \Delta, A\sb1 \quad
  \Gamma \scEnt \Delta, A\sb2}
 {\Gamma \scEnt \Delta, A\sb1 \CAnd A\sb2}
\\[3ex] 
\InfRule{$\COr$-R}
 {\Gamma \scEnt \Delta, A\sb1, A\sb2}
 {\Gamma \scEnt \Delta, A\sb1 \COr A\sb2}
\\[3ex]
\InfRule{$\IImpl$-R}
 {A\sb1, \Gamma \scEnt \Delta, A\sb2}
 {\Gamma \scEnt \Delta, A\sb1 \IImpl A\sb2}
\\[3ex]
\InfRuleC{$\forall$-R}
 {\Gamma \scEnt \Delta, \subst{x \assign y}A}
 {\Gamma \scEnt \Delta, \CForall{x}{A}}
 {\ddag}
\\[3ex]
\InfRule{$\exists$-R}
 {\Gamma \scEnt \Delta, \subst{x \assign t}A}
 {\Gamma \scEnt \Delta, \CExists{x}{A}}
\\[3ex]
\InfRule{$\Not \False$-R}
 {{}}
 {\Gamma \scEnt \Delta, \Not \False}
\\[3ex] 
\InfRule{$\Not \Not$-R}
 {\Gamma \scEnt \Delta, A}
 {\Gamma \scEnt \Delta, \Not \Not A}
\\[3ex] 
\InfRule{$\Not \CAnd$-R}
 {\Gamma \scEnt \Delta, \Not A\sb1, \Not A\sb2}
 {\Gamma \scEnt \Delta, \Not (A\sb1 \CAnd A\sb2)}
\\[3ex] 
\InfRule{$\Not \COr$-R}
 {\Gamma \scEnt \Delta, \Not A\sb1 \quad
  \Gamma \scEnt \Delta, \Not A\sb2}
 {\Gamma \scEnt \Delta, \Not (A\sb1 \COr A\sb2)}
\\[3ex] 
\InfRule{$\Not \IImpl$-R}
 {\Gamma \scEnt \Delta, A\sb1 \quad
  \Gamma \scEnt \Delta, \Not A\sb2}
 {\Gamma \scEnt \Delta, \Not (A\sb1 \IImpl A\sb2)}
\\[3ex]
\InfRule{$\Not \forall$-R}
 {\Gamma \scEnt \Delta, \Not \subst{x \assign t}A}
 {\Gamma \scEnt \Delta, \Not \CForall{x}{A}}
\\[3ex]
\InfRuleC{$\Not \exists$-R}
 {\Gamma \scEnt \Delta, \Not \subst{x \assign y}A}
 {\Gamma \scEnt \Delta, \Not \CExists{x}{A}}
 {\ddag}
\\[3ex]
\InfRuleC{$=$-Repl}
 {\subst{x \assign t\sb1}A, \Gamma \scEnt \Delta}
 {t\sb1 = t\sb2, \subst{x \assign t\sb2}A, \Gamma \scEnt \Delta}
 {$\ast$}
\\[3ex]
\end{tabular}
\end{small}
\\[3ex]
\begin{tabular}{@{}l@{}}
$\ast$ restriction: $A$ is a literal.
\\ 
$\dag$ restriction: $A$ is an atomic formula.
\\
$\ddag$ restriction:
$y$ is not free in $\Gamma$, $y$ is not free in $\Delta$,
$y$ is not free in $A$ unless $x \equiv y$. 
\vspace*{1ex} \par
\end{tabular}
\\
\hline
\end{tabular}
\end{table}
In this table, 
$x$ and $y$ are meta-variables ranging over all variables 
from $\SVar$,
$t$, $t_1$, and $t_2$ are meta-variables ranging over all terms 
from $\STerm{\Sigma}$, 
$A$, $A_1$, and $A_2$ are meta-variables ranging over all formulas 
from $\SForm{\Sigma}$, and
$\Gamma$ and $\Delta$ are meta-variables ranging over all finite 
sets of formulas from $\SForm{\Sigma}$. 

\subsubsection*{Derivations and proofs}

In $\foLPif(\Sigma)$, a \emph{derivation of a sequent 
$\Gamma \scEnt \Delta$ from a finite set of sequents $\mathcal{H}$} is a 
finite sequence $\seq{s_1,\ldots,s_n}$ of sequents such that $s_n$ 
equals $\Gamma \scEnt \Delta$ and, for each $i \in \set{1,\ldots,n}$, 
one of the following conditions holds:
\begin{itemize}
\item
$s_i \in \mathcal{H}$;
\item
$s_i$ is the conclusion of an instance of some inference rule from the 
proof system of $\foLPif(\Sigma)$ whose premises are among 
$s_1,\ldots,s_{i-1}$.
\end{itemize}
A \emph{proof of a sequent $\Gamma \scEnt \Delta$} is a derivation of 
$\Gamma \scEnt \Delta$ from the empty set of sequents.
A sequent $\Gamma \scEnt \Delta$ is said to be \emph{provable} if there 
exists a proof of $\Gamma \scEnt \Delta$.

Let $\Gamma$ and $\Delta$ be sets of formulas from $\SForm{\Sigma}$.
Then $\Delta$ is \emph{derivable} from $\Gamma$, written 
$\Gamma \LDer \Delta$, iff there exist finite sets 
$\Gamma' \subseteq \Gamma$ and $\Delta' \subseteq \Delta$ such that the 
sequent $\Gamma' \scEnt \Delta'$ is provable.

An inference rule that does not belong to the inference rules of some 
proof system is called a \emph{derived inference rule} if there exists 
\pagebreak[2]
a derivation of the conclusion from the premises, using the inference
rules of that proof system, for each instance of the rule.

Let the set $\Gamma_=$ of \emph{equality axioms} be the subset of 
$\SForm{\Sigma}$ consisting of the following formulas:
\begin{itemize}
\item
$\CForall{x}{x = x}$; 
\item
$c = c$ for every $c \in \Func{0}(\Sigma)$;
\item
$\CForall{x_1,y_1,\ldots,x_{n+1},y_{n+1}}
 {\\ \hspace*{2em} x_1 = y_1 \CAnd \ldots \CAnd x_{n+1} = y_{n+1} \IImpl 
   f(x_1,\ldots,x_{n+1}) = f(y_1,\ldots,y_{n+1})}$ \\
for every $f \in \Func{n+1}(\Sigma)$, for every $n \in \Nat$;
\item
$p \IImpl p$ for every $p \in \Pred{0}(\Sigma)$;
\item
$\CForall{x_1,y_1,\ldots,x_{n+1},y_{n+1}}
 {\\ \hspace*{2em} x_1 = y_1 \CAnd \ldots \CAnd x_{n+1} = y_{n+1} \CAnd 
   P(x_1,\ldots,x_{n+1}) \IImpl P(y_1,\ldots,y_{n+1})}$ \\
for every $P \in \Pred{n+1}(\Sigma)$, for every $n \in \Nat$.
\end{itemize}
Then the sequent $\Gamma \scEnt \Delta$ is provable iff
$\Gamma_=, \Gamma \scEnt \Delta$ is provable without using the
inference rules $=$-Refl and $=$-Repl.
This can easily be proved in the same way as Proposition~7.4 
from~\cite{Tak75a} is proved.

In~\cite{Mid20c}, a proof system of \foLPif\ formulated as a 
sequent-style natural deduction system is given.

\subsubsection*{A proof system of $\FOCL(\Sigma)$}

We use the name \FOCL\ here to denote a version of classical logic that 
has the same logical constants, connectives, and quantifiers as \foLPif.

In \FOCL, the same assumptions about symbols are made as in \foLPif\ and
the notion of a signature is defined as in \foLPif.
The languages of $\FOCL(\Sigma)$ and $\foLPif(\Sigma)$ are the same.
A sound and complete sequent calculus proof system of $\FOCL(\Sigma)$ 
can be obtained by adding the following inference rule to the sequent 
calculus proof system of $\foLPif(\Sigma)$:%
\footnote
{If we replace the inference rule $\Not$-R by the inference rule 
 $\Not$-L in the sequent calculus proof system of $\foLPif(\Sigma)$,
 then we obtain a sound and complete proof system of the paracomplete
 analogue of \foLPif.
 The propositional part of that logic (\Klif) is studied
 in e.g.~\cite{Mid17a}.}
\\[1.5ex]
\hspace*{1.5em}
\begin{tabular}{@{}c@{}} 
\InfRule{$\Not$-L}
 {\Gamma \scEnt \Delta, A}
 {\Not A, \Gamma \scEnt \Delta}
\end{tabular}

\section{Truth and Logical Consequence in $\foLPif(\Sigma)$}
\label{INTERPRETATION}

The proof system of $\foLPif(\Sigma)$ is based on the logical 
consequence relation on sets of formulas of $\foLPif(\Sigma)$ defined in 
this section: a sequent $\Gamma \scEnt \Delta$ is provable iff the
logical consequence relation holds between $\Gamma$ and $\Delta$.
This relation is defined in terms of the truth value of formulas of 
$\foLPif(\Sigma)$.
The truth value of formulas is defined relative to a structure and an 
assignment.
First, the notion of a structure and the notion of an assignment are 
introduced.
Next, the truth value of formulas and the logical consequence relation 
on sets of formulas are defined.

\subsubsection*{Structures}

The terms from $\STerm{\Sigma}$ and the formulas from $\SForm{\Sigma}$ 
are interpreted in structures which consist of a non-empty domain of 
individuals and an interpretation of every symbol in the signature $\Sigma$ 
and the equality symbol.
The domain of truth values consists of three values: $\VTrue$ (\emph{true}), 
$\VFalse$ (\emph{false}), and $\VBoth$ (\emph{both true and false}).

A structure $\mathbf{A}$ of $\foLPif(\Sigma)$ consists of:
\begin{itemize}
\item
a set $\mathcal{U}\sp\mathbf{A}$, the \emph{domain of $\mathbf{A}$}, 
such that
$\mathcal{U}\sp\mathbf{A} \neq \emptyset$ and 
$\mathcal{U}\sp\mathbf{A} \inter \set{\VTrue,\VFalse,\VBoth} =
 \emptyset$;
\item
for each $c \in \Func{0}(\Sigma)$, 
\\ \hspace*{20pt}
an element $c\sp\mathbf{A} \in \mathcal{U}\sp\mathbf{A}$;
\vspace*{.5ex}
\item
for each $n \in \Nat$,
for each $f \in \Func{n+1}(\Sigma)$, 
\\ \hspace*{20pt}
a function 
$f\sp\mathbf{A}: 
 {\mathcal{U}\sp\mathbf{A}}^{n+1} \to \mathcal{U}\sp\mathbf{A}$;
\vspace*{.5ex}
\item
for each $p \in \Pred{0}(\Sigma)$, 
\\ \hspace*{20pt}
an element $p\sp\mathbf{A} \in \set{\VTrue,\VFalse,\VBoth}$;
\vspace*{1ex}
\item
for each $n \in \Nat$,
for each $P \in \Pred{n+1}(\Sigma)$, 
\\ \hspace*{20pt}
a function 
$P\sp\mathbf{A}:
 {\mathcal{U}\sp\mathbf{A}}^{n+1} \to \set{\VTrue,\VFalse,\VBoth}$;
\vspace*{.5ex}
\item
a function
$\Meq\sp\mathbf{A}:
 {\mathcal{U}\sp\mathbf{A}}^2 \to \set{\VTrue,\VFalse,\VBoth}$
such that, for each $d \in \mathcal{U}\sp\mathbf{A}$,
\\ \hspace*{20pt}
$\Meq\sp\mathbf{A}(d,d) = \VTrue$ or $\Meq\sp\mathbf{A}(d,d) = \VBoth$.
\end{itemize}
Instead of $w\sp\mathbf{A}$ we write $w$ when it is clear from the 
context that the interpretation of symbol $w$ in structure $\mathbf{A}$ 
is meant.

\subsubsection*{Assignments}

An assignment in a structure $\mathbf{A}$ of $\foLPif(\Sigma)$ assigns 
elements from $\mathcal{U}\sp\mathbf{A}$ to the variables from 
$\SVar$.
The interpretation of the terms from $\STerm{\Sigma}$ and the formulas 
from $\SForm{\Sigma}$ in $\mathbf{A}$ is given with respect to an 
assignment $\alpha$ in $\mathbf{A}$.

Let $\mathbf{A}$ be a structure of $\foLPif(\Sigma)$.
Then an \emph{assignment in $\mathbf{A}$} is a function
$\alpha: \SVar \to \mathcal{U}\sp\mathbf{A}$.
For every assignment $\alpha$ in $\mathbf{A}$, variable 
$x \in \SVar$, and element $d \in \mathcal{U}\sp\mathbf{A}$, we write 
$\alpha(x \to d)$ for the assignment $\alpha'$ in $\mathbf{A}$ such that 
$\alpha'(x) = d$ and $\alpha'(y) = \alpha(y)$ if $y \not\equiv x$.

\subsubsection*{Valuations and models}

The valuation of the terms from $\STerm{\Sigma}$ is given by a 
function mapping term $t$, structure $\mathbf{A}$ and assignment 
$\alpha$ in $\mathbf{A}$ to the element of $\mathcal{U}\sp\mathbf{A}$ 
that is the value of $t$ in $\mathbf{A}$ under assignment $\alpha$.
Similarly, the valuation of the formulas from $\SForm{\Sigma}$ is 
given by a function mapping formula $A$, structure $\mathbf{A}$ and 
assignment $\alpha$ in $\mathbf{A}$ to the element of 
$\set{\VTrue,\VFalse,\VBoth}$ that is the truth value of $A$ in 
$\mathbf{A}$ under assignment $\alpha$.
We write $\Term{t}{\mathbf{A}}{\alpha}$ and 
$\Term{A}{\mathbf{A}}{\alpha}$ for these valuations.

The valuation functions for the terms from $\STerm{\Sigma}$ and the 
formulas from $\SForm{\Sigma}$ are inductively defined in 
Table~\ref{table-interpretation}.
\begin{table}[t!]
\caption{Valuations of terms and formulas of $\foLPif(\Sigma)$}
\label{table-interpretation}
\vspace*{-3ex} \par \mbox{} 
\renewcommand{\arraystretch}{1.25}
\centering
\begin{array}[t]{rcl}
\hline
\mbox{} \\[-2.5ex]
\Term{x}{\mathbf{A}}{\alpha} & = &
 \begin{array}[t]{l}
 \alpha(x) \;,
 \end{array}
\\[.5ex]
\Term{c}{\mathbf{A}}{\alpha} & = &
 \begin{array}[t]{l}
 c\sp\mathbf{A} \;,
 \end{array}
\\[.5ex]
\Term{f(t\sb1,\ldots,t\sb{n+1})}{\mathbf{A}}{\alpha} & = &
 \begin{array}[t]{l}
 f\sp\mathbf{A}(\Term{t\sb1}{\mathbf{A}}{\alpha},\ldots,
                \Term{t\sb{n+1}}{\mathbf{A}}{\alpha})
 \end{array}
\\[1.5ex]
\Term{\False}{\mathbf{A}}{\alpha} & = &
 \begin{array}[t]{l}
 \VFalse \;,
 \end{array}
\\[.5ex]
\Term{p}{\mathbf{A}}{\alpha} & = &
 \begin{array}[t]{l}
 p\sp\mathbf{A} \;,
 \end{array}
\\[.5ex]
\Term{P(t\sb1,\ldots,t\sb{n+1})}{\mathbf{A}}{\alpha} & = &
 \begin{array}[t]{l}
 P\sp\mathbf{A}(\Term{t\sb1}{\mathbf{A}}{\alpha},\ldots,
                \Term{t\sb{n+1}}{\mathbf{A}}{\alpha}) \;,
 \end{array}
\\[.5ex]
\Term{t\sb1 = t\sb2}{\mathbf{A}}{\alpha} & = &
 \begin{array}[t]{l}
 \Meq\sp\mathbf{A}(\Term{t\sb1}{\mathbf{A}}{\alpha},
                   \Term{t\sb2}{\mathbf{A}}{\alpha}) \;,
\end{array}
\\[1.5ex]
\Term{\Not A}{\mathbf{A}}{\alpha} & = &
 \left \{
 \begin{array}{l@{\;\;}l}
 \VTrue  & \mathrm{if}\; \Term{A}{\mathbf{A}}{\alpha} = \VFalse \\
 \VFalse & \mathrm{if}\; \Term{A}{\mathbf{A}}{\alpha} = \VTrue \\
 \VBoth  & \mathrm{otherwise},
 \end{array}
 \right.
\vspace*{.5ex} \\
\Term{A\sb1 \CAnd A\sb2}{\mathbf{A}}{\alpha} & = &
 \left \{
 \begin{array}{l@{\;\;}l}
 \VTrue  & \mathrm{if}\; 
           \Term{A\sb1}{\mathbf{A}}{\alpha} = \VTrue\; \mathrm{and}\;
           \Term{A\sb2}{\mathbf{A}}{\alpha} = \VTrue \\
 \VFalse & \mathrm{if}\;
           \Term{A\sb1}{\mathbf{A}}{\alpha} = \VFalse\; \mathrm{or}\;
           \Term{A\sb2}{\mathbf{A}}{\alpha} = \VFalse \\
 \VBoth  & \mathrm{otherwise},
 \end{array}
 \right.
\vspace*{.5ex} \\
\Term{A\sb1 \COr A\sb2}{\mathbf{A}}{\alpha} & = &
 \left \{
 \begin{array}{l@{\;\;}l}
 \VTrue  & \mathrm{if}\; 
           \Term{A\sb1}{\mathbf{A}}{\alpha} = \VTrue\; \mathrm{or}\;
           \Term{A\sb2}{\mathbf{A}}{\alpha} = \VTrue \\
 \VFalse & \mathrm{if}\;
           \Term{A\sb1}{\mathbf{A}}{\alpha} = \VFalse\; \mathrm{and}\;
           \Term{A\sb2}{\mathbf{A}}{\alpha} = \VFalse \\
 \VBoth  & \mathrm{otherwise},
 \end{array}
 \right.
\vspace*{.5ex} \\
\Term{A\sb1 \IImpl A\sb2}{\mathbf{A}}{\alpha} & = &
 \left \{
 \begin{array}{l@{\;\;}l}
 \VTrue  & \mathrm{if}\; 
           \Term{A\sb1}{\mathbf{A}}{\alpha} = \VFalse\; \mathrm{or}\;
           \Term{A\sb2}{\mathbf{A}}{\alpha} = \VTrue \\
 \VFalse & \mathrm{if}\;
           \Term{A\sb1}{\mathbf{A}}{\alpha} \neq \VFalse\; \mathrm{and}\;
           \Term{A\sb2}{\mathbf{A}}{\alpha} = \VFalse \\
 \VBoth  & \mathrm{otherwise},
 \end{array}
 \right.
\vspace*{.5ex} \\
\Term{\CForall{x}{A}}{\mathbf{A}}{\alpha} & = &
 \left \{
 \begin{array}{l@{\;\;}l}
 \VTrue  & \mathrm{if},\;
           \mathrm{for\; all}\;  d \in \mathcal{U}\sp\mathbf{A},\;
           \Term{A}{\mathbf{A}}{\alpha(x \to d)} = \VTrue \\
 \VFalse & \mathrm{if},\;
           \mathrm{for\; some}\; d \in \mathcal{U}\sp\mathbf{A},\;
           \Term{A}{\mathbf{A}}{\alpha(x \to d)} = \VFalse \\
 \VBoth  & \mathrm{otherwise}.
 \end{array}
 \right.
\vspace*{.5ex} \\
\Term{\CExists{x}{A}}{\mathbf{A}}{\alpha} & = &
 \left \{
 \begin{array}{l@{\;\;}l}
 \VTrue  & \mathrm{if},\;
           \mathrm{for\; some}\;  d \in \mathcal{U}\sp\mathbf{A},\;
           \Term{A}{\mathbf{A}}{\alpha(x \to d)} = \VTrue \\
 \VFalse & \mathrm{if},\;
           \mathrm{for\; all}\; d \in \mathcal{U}\sp\mathbf{A},\;
           \Term{A}{\mathbf{A}}{\alpha(x \to d)} = \VFalse \\
 \VBoth  & \mathrm{otherwise}.
 \end{array}
 \right.
\vspace*{1.25ex} \\
\hline
\end{array}
\vspace*{-1.5ex}
\end{table}
In this table, 
$x$ is a meta-variable ranging over all variables from 
$\SVar$, 
$c$ is a meta-variable ranging over all function symbols from 
$\Func{0}(\Sigma)$,
$f$ is a meta-variable ranging over all function symbols from
$\Func{n+1}(\Sigma)$ (where $n$ is understood from the context), 
$p$ is a meta-variable ranging over all predicate symbols from 
$\Pred{0}(\Sigma)$,
$P$ is a meta-variable ranging over all  predicate symbols from
$\Pred{n+1}(\Sigma)$ (where $n$ is understood from the context), 
$t_1$, \ldots, $t_{n+1}$ are meta-variables ranging over all terms 
from $\STerm{\Sigma}$, and 
$A$, $A_1$, and $A_2$ are meta-variables ranging over all formulas 
from~$\SForm{\Sigma}$.

The following theorem is a decidability result concerning valuations of
formulas in structures with a finite domain. 
\begin{theorem}
\label{theorem-decidable}
Let $\mathbf{A}$ be a structure of $\foLPif(\Sigma)$ such that
$\mathcal{U}\sp\mathbf{A}$ is finite, and
let $\alpha$ be an assignment in $\mathbf{A}$.
Then, for all $A \in \SForm{\Sigma}$,\,  
$\Term{A}{\mathbf{A}}{\alpha} \in \set{\VTrue,\VBoth}$ is decidable.
\end{theorem}
\begin{proof}
This is easy to prove by induction on the structure of $A$.
\qed
\end{proof}

Let $\Gamma$ be a set of formulas from $\SForm{\Sigma}$.
Then a \emph{model of $\Gamma$} is a structure~$\mathbf{A}$ of 
$\foLPif(\Sigma)$ such that, for all assignments $\alpha$ in 
$\mathbf{A}$, for all $A \in \Gamma$, 
$\Term{A}{\mathbf{A}}{\alpha} \in \set{\VTrue,\VBoth}$.

\subsubsection*{Logical consequence}

Let $\Gamma$ and $\Delta$ be sets of formulas from $\SForm{\Sigma}$.
Then \emph{$\Delta$ is a logical consequence of $\Gamma$}, written 
$\Gamma \LCon \Delta$, iff
for all structures $\mathbf{A}$ of $\foLPif(\Sigma)$,
for all assignments $\alpha$ in $\mathbf{A}$,
$\Term{A}{\mathbf{A}}{\alpha} = \VFalse$ for some $A \in \Gamma$ or
$\Term{A'}{\mathbf{A}}{\alpha} \in \set{\VTrue,\VBoth}$ for some 
$A' \in \Delta$.

The sequent calculus proof system of $\foLPif(\Sigma)$ presented in 
Section~\ref{PROOF-RULES} is sound and complete with respect to logical 
consequence as defined above.
\begin{theorem}
\label{theorem-sound-complete}
Let $\Gamma$ and $\Delta$ be finite sets of formulas from 
$\SForm{\Sigma}$. 
Then $\Gamma \LDer \Delta$ iff $\Gamma \LCon \Delta$.
\end{theorem}
\begin{proof}
In the proof of this theorem use is made of the fact that a sound and 
complete sequent calculus proof system for \foLPc, a logic similar to 
\foLPif, is available in~\cite{Pic18a}.
The differences between the two logics are:
\begin{itemize}
\item
the proof system of \foLPif\ does not include a cut rule and the proof 
system of \foLPc\ includes a cut rule, but the latter proof system has 
the cut-elimination property;
\item
the $\Not$-R rule and the Repl rule from the proof system of \foLPif\
differ from the $\Not$-R rule and the Repl rule from the proof system of 
\foLPc, but in either proof systems the $\Not$-R rule and the Repl rule 
from the other proof system are derived inference rules;
\item
the logical symbols of \foLPc\ include the \emph{consistency} connective 
$\circ$ and the logical symbols of \foLPif\ do not include this logical
symbol, but formulas with it as outermost operator can be defined as  
abbreviations of formulas in \foLPif\ as follows:
$\circ A$ stands for $(A \IImpl \False) \COr (\Not A \IImpl \False)$;
\item
the logical symbols of \foLPif\ include $\False$, $\IImpl$, $\CAnd$, and
$\forall$ and the logical symbols of \foLPc\ do not include these 
logical symbols, but formulas with them as outermost operator can be 
defined as abbreviations of formulas in \foLPc\ as follows:
$\False$~stands for $A \CAnd \Not A \CAnd \circ A$ where $A$ is an 
arbitrary atomic formula,
$A_1 \IImpl A_2$ stands for $(\Not A_1 \CAnd \circ A_1) \COr A_2$,
$A_1 \CAnd A_2$ stands for $\Not (\Not A_1 \COr \Not A_2)$, and
$\CForall{x}{A}$ stands for $\Not \CExists{x}{\Not A}$.
\end{itemize}
For each formula of one of the two logics which is defined above as an 
abbreviation of a formula in the other logic, the valuation of the 
former formula in the former logic is the same as the valuation of 
the latter formula in the latter logic.
Moreover, the first two differences mentioned above have no effect on 
the sequents that can be proved.
Therefore, the sequent calculus proof system of \foLPif\ is sound and 
complete if, for each logical symbol missing in one of the logics, the 
inference rules for that symbol in the proof system of the other logic 
become derived inference rules in the proof system of the former logic 
when the formulas with that symbol as outermost operator are taken for 
abbreviations of formulas as defined above.  
It is a routine matter to prove this.
\qed
\end{proof}
A non-standard, indirect proof of soundness and completeness is outlined 
above. 
This proof outline clarifies why \foLPif\ is called `essentially the 
same as' \foLPc\ in Section~\ref{sect-intro}. 
Moreover, it follows from this proof outline that the admissibility of 
the structural inference rules of cut and weakening in \foLPc\ carries 
over to~\foLPif.
A direct proof of soundness and completeness can be given along the same 
lines as in the proof of Theorem~1 from~\cite{Mid23b}.

There are two minor differences between \foLPif\ and \foLPc\ that are 
not mentioned in the proof outline above.
The first difference is that a predicate symbol is interpreted in 
\foLPc\ as what is sometimes called a paraconsistent relation 
(see e.g.~\cite{BS95a}) and in \foLPif\ as what may be called the 
characteristic function of such a relation.
However, this difference is nullified in the valuation of formulas.
The second difference is that in \foLPc\ signatures are restricted to 
signatures $\Sigma$ for which $\Pred{0}(\Sigma) = \emptyset$. 
By consulting the soundness and completeness proofs in~\cite{Pic18a}, it
becomes immediately clear that, as expected, this restriction can be 
removed without effect on the soundness and completeness.

\subsubsection*{Abbreviations}

From Section~\ref{DATABASE} on, we use $\Cons A$ and $A_1 \SImpl A_2$ as 
abbreviations for formulas in \foLPif.
These abbreviations are defined as follows: 
$\Cons A$ stands for $(A \IImpl \False) \COr (\Not A \IImpl \False)$ and
$A_1 \SImpl A_2$ stands for 
$(A_1 \IImpl A_2) \CAnd (\Not A_2 \IImpl \Not A_1)$.
It follows from these definitions that:
\[
\begin{array}[t]{l@{\;\;}c@{\;\;}l}
\Term{\Cons A}{\mathbf{A}}{\alpha} & = &
 \left \{
 \begin{array}{l@{\;\;}l}
 \VTrue  & \mathrm{if}\; 
           \Term{A}{\mathbf{A}}{\alpha} = \VTrue\; \mathrm{or}\;
           \Term{A}{\mathbf{A}}{\alpha} = \VFalse \\
 \VFalse & \mathrm{otherwise},
 \end{array}
 \right.
\vspace*{.5ex} \\
\Term{A\sb1 \SImpl A\sb2}{\mathbf{A}}{\alpha} & = &
 \left \{
 \begin{array}{l@{\;\;}l}
 \VTrue  & \mathrm{if}\; 
           \Term{A\sb1}{\mathbf{A}}{\alpha} = \VFalse\; \mathrm{or}\;
           \Term{A\sb2}{\mathbf{A}}{\alpha} = \VTrue \\
 \VBoth  & \mathrm{if}\;
           \Term{A\sb1}{\mathbf{A}}{\alpha} = \VBoth\; \mathrm{and}\;
           \Term{A\sb2}{\mathbf{A}}{\alpha} = \VBoth \\
 \VFalse & \mathrm{otherwise},
 \end{array}
 \right.
\end{array}
\]

\section{Relational Databases Viewed through \foLPif}
\label{DATABASE}

\sloppy
In this section, relational databases are considered from the 
perspective of \foLPif.
A relational database can be considered from a logical point of view in 
two different ways: either as a model of a logical theory 
(the model-theoretic view) or as a logical theory 
(the proof-theoretic view). 
Here, the second viewpoint is taken.
In the definition of the notion of a relational database, use is made of
the notions of a relational language and a relational theory.
The latter two notions are defined first.
The definitions given in this section are based on those given 
in~\cite{Rei84a}.
However, types are ignored for the sake of 
simplicity (cf.~\cite{GMN84a,Var86a}). 

\subsubsection*{Relational languages}

The pair $(\Sigma,\SForm{\Sigma})$, where $\Sigma$ is a signature, is 
called the \emph{language of $\foLPif(\Sigma)$}.
If $\Sigma$ satisfies particular conditions, then the language of 
$\foLPif(\Sigma)$ is considered a relational language.

Let $\Sigma$ be a signature.
Then the language $R = (\Sigma,\SForm{\Sigma})$ of $\foLPif(\Sigma)$ is 
a \emph{relational language} iff it satisfies the following conditions:
\begin{itemize} 
\item
$\Func{0}(\Sigma)$ is non-empty and finite;
\item
$\Union \set{\Func{n+1}(\Sigma) \where n \in \Nat}$ is empty;\vsp
\item
$\Pred{0}(\Sigma)$ is empty;\vsp
\item
$\Union \set{\Pred{n+1}(\Sigma) \where n \in \Nat}$ is finite.
\end{itemize}

\subsubsection*{Relational theories}

Below, we will introduce the notion of a relational theory.
In the definition of a relational theory, use is made of four 
auxiliary notions, namely the notions of an atomic fact, a domain 
closure axiom, a unique name axiom set, and a completion axiom.
These auxiliary notions are defined first.

Let $R = (\Sigma,\SForm{\Sigma})$ be a relational language.
Then an \emph{atomic fact for $R$} is a formula from $\SForm{\Sigma}$ of 
the form $P(c_1,\ldots,c_{n+1})$, where $P \in \Pred{n+1}(\Sigma)$ and 
$c_1,\ldots,c_{n+1} \in \Func{0}(\Sigma)$. 

Let $R = (\Sigma,\SForm{\Sigma})$ be a relational language.
Then the \emph{equality consistency axiom for $R$} is the formula
\[
\CForall{x,x'}{\Cons (x = x')}\;.
\]

Let $R = (\Sigma,\SForm{\Sigma})$ be a relational language and
let $c_1,\ldots,c_n$ be all members of $\Func{0}(\Sigma)$.
Then the \emph{domain closure axiom for $R$} is the formula
\[
\CForall{x}{(x = c_1 \COr \ldots \COr x = c_n)}
\]
and the \emph{unique name axiom set for $R$} is the set of formulas
\[\set{\Not (c_i = c_j) \where 1 \leq i < j \leq n}\;.
\]

Let $R = (\Sigma,\SForm{\Sigma})$ be a relational language,
let $\Lambda \subseteq \SForm{\Sigma}$ be a finite set of atomic facts
for $R$, and let $P \in \Pred{n+1}(\Sigma)$ ($n \in \Nat$). 
Suppose that there exist formulas in $\Lambda$ in which $P$ occurs and
let $P(c^1_1,\ldots,c^1_{n+1}),\ldots,P(c^m_1,\ldots,c^m_{n+1})$ be 
all formulas from $\Lambda$ in which $P$ occurs.
Then the \emph{$P$-completion axiom for $\Lambda$} is the formula
\[
\begin{array}[t]{@{}l@{}}
\CForall{x_1,\ldots,x_{n+1}}
 {P(x_1,\ldots,x_{n+1}) \SImpl {} \\ \hspace*{2em} 
  x_1 = c^1_1 \CAnd \ldots \CAnd x_{n+1} = c^1_{n+1}
   \,\COr\, \ldots \,\COr\,
  x_1 = c^m_1 \CAnd \ldots \CAnd x_{n+1} = c^m_{n+1}}\;.
\end{array}
\]
Suppose that there does not exist a formula in $\Lambda$ in which $P$ 
occurs.
Then the \emph{$P$-completion axiom for $\Lambda$} is the formula
\[
\begin{array}[t]{@{}l@{}}
\CForall{x_1,\ldots,x_{n+1}}{P(x_1,\ldots,x_{n+1}) \SImpl \False}\;.
\end{array}
\]

Let $R = (\Sigma,\SForm{\Sigma})$ be a relational language.
Then the \emph{relational structure axioms for $R$}, written
$\mathit{RSA}(R)$, is the set of all formulas 
$A \in \SForm{\Sigma}$ for which one of the following holds:
\begin{itemize} 
\item
$A$ is the equality consistency axiom for $R$;
\item
$A$ is the domain closure axiom for $R$;
\item
$A$ is an element of the unique name axiom set for $R$.
\end{itemize} 

Let $R = (\Sigma,\SForm{\Sigma})$ be a relational language, and
let $\Lambda \subseteq \SForm{\Sigma}$ be a finite set of atomic facts
for $R$.
Then the \emph{relational theory for $R$ with basis $\Lambda$}, written
$\mathit{RT}(R,\Lambda)$, is the set of all formulas 
$A \in \SForm{\Sigma}$ for which one of the following holds:
\pagebreak[2]
\begin{itemize} 
\item
$A \in \mathit{RSA}(R)$;
\item
$A \in \Lambda$;
\item
$A$ is the $P$-completion axiom for $\Lambda$ for some 
$P \in \Union \set{\Pred{n+1}(\Sigma) \where n \in \Nat}$.
\end{itemize} 
A set $\Theta \subseteq \SForm{\Sigma}$ is called a 
\emph{relational theory for $R$} if $\Theta = \mathit{RT}(R,\Lambda)$ 
for some finite set $\Lambda \subseteq \SForm{\Sigma}$ of atomic facts 
for $R$.
The elements of this unique $\Lambda$ are called the 
\emph{atomic facts of $\Theta$}.

The following theorem is a decidability result concerning provability 
of sequents $\Gamma \scEnt A$ where $\Gamma$ includes the relational 
structure axioms for some relational language.
\begin{theorem}
\label{theorem-decidable-add}
Let $R = (\Sigma,\SForm{\Sigma})$ be a relational language, and
let $\Gamma$ be a finite subset of $\SForm{\Sigma}$ such that 
$\mathit{RSA}(R) \subseteq \Gamma$.
Then it is decidable whether, for a formula $A \in \SForm{\Sigma}$,\, 
$\Gamma \scEnt A$ is provable. 
\end{theorem}
\begin{proof}
Because $\mathit{RSA}(R) \subseteq \Gamma$, it is sufficient to consider 
only structures that are models of $\mathit{RSA}(R)$.
The domains of these structures have the same finite cardinality. 
Because in addition there are finitely many predicate symbols in 
$\Sigma$, there exist moreover only finitely many of these structures.

Clearly, it is sufficient to consider only the restrictions of 
assignments to the set of all variables that occur free in 
$\Gamma \union \set{A}$.
Because the set of all variables that occur free in 
$\Gamma \union \set{A}$ is finite and the domain of the structures to be 
considered is finite, there exist only finitely many such restrictions 
and those restrictions are finite.

It follows easily from the above-mentioned finiteness properties and 
Theorems~\ref{theorem-decidable} and~\ref{theorem-sound-complete} that 
it is decidable whether,
for a formula $A \in \SForm{\Sigma}$,\, $\Gamma \scEnt A$ is provable.
\qed
\end{proof}

\subsubsection*{Relational databases}

Having defined the notions of an relational language and a relational 
theory, we are ready to define the notion of a relational database in 
the setting of \foLPif.

A \emph{relational database} $\mathit{DB}$ is a triple 
$(R,\Theta,\Xi)$, where:
\begin{itemize} 
\item
$R = (\Sigma,\SForm{\Sigma})$ is a relational language;
\item
$\Theta$ is a relational theory for $R$;
\item
$\Xi$ is a finite subset of $\SForm{\Sigma}$.  
\end{itemize} 
$\Theta$ is called the \emph{relational theory of $\mathit{DB}$} and
$\Xi$ is called the \emph{set of integrity constraints of 
$\mathit{DB}$}.

The set $\Xi$ of integrity constraints of a relational database 
$\mathit{DB} = (R,\Theta,\Xi)$ can be seen as a set of assumptions about 
the relational theory of the relational database $\Theta$.
If the relational theory agrees with these assumptions, then the
relational database is called consistent.

Let $R = (\Sigma,\SForm{\Sigma})$ be a relational language, and
let $\mathit{DB} = (R,\Theta,\Xi)$ be a relational database.
Then $\mathit{DB}$ \emph{is consistent} iff, 
for each $A \in \SForm{\Sigma}$ such that $A$ is an atomic fact for $R$
or $A$ is of the form $\Not A'$ where $A'$ is an atomic fact for~$R$:
\[
\renewcommand{\arraystretch}{1.35}
\begin{tabular}{c} 
$\Theta \scEnt A$ is provable only if $\Theta,\Xi \scEnt \Cons A$ is 
provable.
\end{tabular}
\]

Notice that, if $\mathit{DB}$ is not consistent, $\Theta,\Xi \scEnt A$
is provable with the sequent calculus proof system of $\FOCL(\Sigma)$ 
for all $A \in \SForm{\Sigma}$.
However, the sequent calculus proof system of $\foLPif(\Sigma)$ rules 
out such an explosion.

\subsubsection*{Models of relational theories}

The models of relational theories for a relational language 
$R = (\Sigma,\SForm{\Sigma})$ are structures of $\foLPif(\Sigma)$ of 
a special kind.

Let $R = (\Sigma,\SForm{\Sigma})$ be a relational language.
Then a \emph{relational structure for $R$} is a structure $\mathbf{A}$ 
of $\foLPif(\Sigma)$ such that:
\begin{itemize} 
\item
for all $d_1,d_2 \in \mathcal{U}\sp\mathbf{A}$, 
${=\sp\mathbf{A}}(d_1,d_2) \in \set{\VTrue,\VFalse}$;
\item
for all $d \in \mathcal{U}\sp\mathbf{A}$, 
there exists a $c \in \Func{0}(\Sigma)$ such that 
${=\sp\mathbf{A}}(d,c\sp\mathbf{A}) = \VTrue$;
\item
for all $c_1,c_2 \in\Func{0}(\Sigma)$, 
${=\sp\mathbf{A}}({c_1}\sp\mathbf{A},{c_2}\sp\mathbf{A}) = \VTrue$ only 
if $c_1 \equiv c_2$.
\end{itemize} 

Let $R = (\Sigma,\SForm{\Sigma})$ be a relational language, and
let $\mathbf{A}$ be a structure of $\foLPif(\Sigma)$.
Then $\mathbf{A}$ is a relational structure for $R$ iff,
for all assignments $\alpha$ in $\mathbf{A}$,
for all $A \in \mathit{RSA}(R)$,\,
$\Term{A}{\mathbf{A}}{\alpha} \in \set{\VTrue,\VBoth}$.
Moreover, let $\Theta$ be a relational theory for $R$.
Then all models of $\Theta$ are relational structures for $R$ because 
$\mathit{RSA}(R) \subseteq \Theta$.\,
$\Theta$ does not have a unique model up to isomorphism.
$\Theta$'s predicate completion axioms fail to enforce a unique model up 
to isomorphism.
However, identification of $\VTrue$ and $\VBoth$ in the models of 
$\Theta$ yields uniqueness up to isomorphism. 

Let $R = (\Sigma,\SForm{\Sigma})$ be a relational language, and
let $\mathbf{A}$ be a relational structure for $R$.
Then we write $\nabla \mathbf{A}$ for the relational structure 
$\mathbf{A}'$ for $R$ such that:
\begin{itemize}
\item
$\mathcal{U}\sp{\mathbf{A}'} = \mathcal{U}\sp\mathbf{A}$;
\item
for each $c\in \Func{0}(\Sigma)$, 
$c\sp{\mathbf{A}'} = c\sp\mathbf{A}$; 
\item
for each $n \in \Nat$,
for each $P \in \Pred{n+1}(\Sigma)$, 
for each $d_1,\ldots,d_{n+1} \in \mathcal{U}\sp{\mathbf{A}'}$,
$P\sp{\mathbf{A}'}(d_1,\ldots,d_{n+1}) = 
 \left \{
 \begin{array}{l@{\;\;}l}
 \VTrue  & \mathrm{if}\; P\sp\mathbf{A}(d_1,\ldots,d_{n+1}) \in
                         \set{\VTrue,\VBoth} \\
 \VFalse & \mathrm{otherwise};
 \end{array}
 \right.$
\item
for each $d_1,d_2 \in \mathcal{U}\sp{\mathbf{A}'}$,
${=\sp{\mathbf{A}'}}(d_1,d_2) \,=\, {=\sp{\mathbf{A}}}(d_1,d_2)$.
\end{itemize}

Let $R = (\Sigma,\SForm{\Sigma})$ be a relational language, 
let $\Theta$ be a relational theory for $R$, and
let $\mathbf{A}$ be a model of $\Theta$.
Then $\nabla \mathbf{A}$, i.e.\ $\mathbf{A}$ with $\VTrue$ and $\VBoth$ 
identified, is in essence a relational database as originally introduced 
in~\cite{Cod70a}.

\begin{theorem}
\label{theorem-unique-rmodel}
Let $R = (\Sigma,\SForm{\Sigma})$ be a relational language, 
let $\Theta$ be a relational theory for $R$, and
let $\mathbf{A}$ and $\mathbf{A}'$ be models of $\Theta$.
Then $\nabla \mathbf{A}$ and $\nabla \mathbf{A}'$ are isomorphic 
relational structures.
\end{theorem}
\begin{proof}
The proof goes in almost the same way as the proof of part~1 of 
Theorem~3.1 from~\cite{Rei84a}.
The only point of attention is that it may be the case that, for some 
$P \in \Pred{n+1}(\Sigma)$ and 
$c_1,\ldots,c_{n+1} \in \Func{0}(\Sigma)$ ($n \in \Nat$), 
either $\Term{P(c_1,\ldots,c_{n+1})}{\mathbf{A}}{\alpha} = \VTrue$ and
$\Term{P(c_1,\ldots,c_{n+1})}{\mathbf{A'}}{\alpha} = \VBoth$ or
$\Term{P(c_1,\ldots,c_{n+1})}{\mathbf{A}}{\alpha} = \VBoth$ and
$\Term{P(c_1,\ldots,c_{n+1})}{\mathbf{A'}}{\alpha} = \VTrue$.
But, if this is the case, 
$\Term{P(c_1,\ldots,c_{n+1})}{\mathbf{\nabla A}}{\alpha} = \VTrue$ and
$\Term{P(c_1,\ldots,c_{n+1})}{\mathbf{\nabla A'}}{\alpha} = \VTrue$.
\qed
\end{proof}

\begin{theorem}
\label{theorem-exists-rtheory}
Let $R = (\Sigma,\SForm{\Sigma})$ be a relational language, and
let $\mathbf{A}$ be a relational structure for $R$.
Then there exists a relational theory $\Theta$ for $R$ such that 
$\mathbf{A}$ is a model of $\Theta$.
\end{theorem}
\begin{proof}
The proof goes in the same way as the proof of part~2 of Theorem~3.1 
from~\cite{Rei84a}.
\qed
\end{proof}

\section{Query Answering Viewed through \foLPif}
\label{QUERY-ANSWERING}

In this section, queries applicable to a relational database and their
answers are considered from the perspective of \foLPif.
As a matter of fact, the queries introduced below are closely related to 
the relational-calculus-oriented queries originally originally 
introduced in~\cite{Cod72a}.

\subsubsection*{Queries}

As to be expected in the current setting, a query applicable to a 
relational database involves a formula of \foLPif.

Let $R = (\Sigma,\SForm{\Sigma})$ be a relational language.
Then a \emph{query for $R$} is an expression of the form
$(x_1,\ldots,x_n) \suchthat A$, where:
\begin{itemize} 
\item
$x_1,\ldots,x_n \in \SVar$;
\item
$A \in \SForm{\Sigma}$ and all variables that are free in $A$ are among
$x_1,\ldots,x_n$.
\end{itemize} 

Let $\mathit{DB} = (R,\Theta,\Xi)$ be a relational database.
Then a query is \emph{applicable to $\mathit{DB}$} iff it is a query for
$R$.

\subsubsection*{Answers}

Answering a query with respect to a consistent relational database 
amounts to looking for closed instances of the formula concerned that 
are logical consequences of a relational theory.
The main issue concerning query answering is how to deal with 
inconsistent relational databases.

Let $R = (\Sigma,\SForm{\Sigma})$ be a relational language, 
let $\mathit{DB} = (R,\Theta,\Xi)$ be a relational database, and
let $(x_1,\ldots,x_n) \suchthat A$ be a query that is applicable to 
$\mathit{DB}$.
Then an \emph{answer to $(x_1,\ldots,x_n) \suchthat A$ with respect to 
$\mathit{DB}$} is a $(c_1,\ldots,c_n) \in {\Func{0}(\Sigma)}^n$ for 
which 
$\Theta \scEnt \subst{x_1 \assign c_1}\ldots\subst{x_n \assign c_n}A$
is provable.

The above definition of an answer to a query with respect to a database
does not take into account the integrity constraints of the database
concerned.

\subsubsection*{Consistent answers}

The definition of a consistent answer given below is based on the 
following:
\begin{itemize}
\item
the observation that the formula that corresponds to an answer, being a 
logical consequence of the relational theory of the database, is also a
logical consequence of one or more sets of atomic facts and negations of 
atomic facts that are logical consequences of the relational theory of 
the database;
\item
the idea that in the case of a consistent answer there must be such a 
set that does not contain an atomic fact or negation of an atomic fact 
that causes the database to be inconsistent.
\end{itemize}

Let $R = (\Sigma,\SForm{\Sigma})$ be a relational language.
Then an \emph{semi-atomic fact for $R$} is a formula from 
$\SForm{\Sigma}$ of the form $P(c_1,\ldots,c_{n+1})$ or the form 
$\Not P(c_1,\ldots,c_{n+1})$, where $P \in \Pred{n+1}(\Sigma)$ and 
$c_1,\ldots,c_{n+1} \in \Func{0}(\Sigma)$. 

Let $R = (\Sigma,\SForm{\Sigma})$ be a relational language, 
let $\mathit{DB} = (R,\Theta,\Xi)$ be a relational database, and
let $(x_1,\ldots,x_n) \suchthat A$ be a query that is applicable to 
$\mathit{DB}$. 
Then a \emph{consistent answer to $(x_1,\ldots,x_n) \suchthat A$ with 
respect to $\mathit{DB}$} is a
$(c_1,\ldots,c_n) \in {\Func{0}(\Sigma)}^n$ 
for which there exists a
$\Phi \subseteq
 \set{A' \where A' \mbox{ is a semi-atomic fact for $R$}}$ such that:
\begin{itemize}
\item
for all $A' \in \Phi$,\, $\Theta \scEnt A'$ is provable and
$\Theta, \Xi \scEnt \Cons A'$ is provable; 
\item
$\Phi, \mathit{RSA}(R) \scEnt
 \subst{x_1 \assign c_1}\ldots\subst{x_n \assign c_n}A$
is provable. 
\end{itemize}

The above definition of a consistent answer to a query with respect to 
a database is reminiscent of the definition of a consistent answer to a 
query with respect to a database given in~\cite{Bry97a}.
It simply accepts that a database is inconsistent and excludes the 
source or sources of the inconsistency from being used in consistent
query answering.

\subsubsection*{Strongly consistent answers}

The definition of a strongly consistent answer given below is not so
tolerant of inconsistency and makes use of consistent repairs of the 
database.
The idea is that an answer is strongly consistent if it is an answer 
with respect to any minimally repaired version of the original database.

Let $R = (\Sigma,\SForm{\Sigma})$ be a relational language, and
let $\Lambda \subseteq \SForm{\Sigma}$ be a finite set of atomic facts
for $R$.
Then, following~\cite{ABC99a}, the binary relation $\leq_\Lambda$ on the 
set of all finite sets of atomic facts for $R$ is defined by:
\[\Lambda' \leq_\Lambda \Lambda'' \mbox{ iff }
  (\Lambda \diff \Lambda') \union (\Lambda' \diff \Lambda) \subseteq
  (\Lambda \diff \Lambda'') \union (\Lambda'' \diff \Lambda)\;.
\]
Intuitively, $\Lambda' \leq_\Lambda \Lambda''$ indicates that the extent
to which $\Lambda'$ differs from $\Lambda$ is less than the extent to 
which $\Lambda''$ differs from $\Lambda$.

Let $R = (\Sigma,\SForm{\Sigma})$ be a relational language, 
let $\Lambda \subseteq \SForm{\Sigma}$ be a finite set of atomic facts
for $R$, and
let $\Xi$ is a finite subset of $\SForm{\Sigma}$. 
Then \emph{$\Lambda$ is consistent with $\Xi$} iff
for all semi-atomic facts $A$ for $R$,
$\mathit{RT}(R,\Lambda) \scEnt A$ is provable only if 
$\mathit{RT}(R,\Lambda),\Xi \scEnt \Not A$ is not provable.
We write $\mathit{Con}(\Xi)$ for the set of all finite sets of atomic 
facts for $R$ that are consistent with $\Xi$.

Let $R = (\Sigma,\SForm{\Sigma})$ be a relational language, 
let $\Lambda \subseteq \SForm{\Sigma}$ be a finite set of atomic facts
for $R$,
let $\mathit{DB} = (R,\mathit{RT}(R,\Lambda),\Xi)$ be a relational 
database, and
let $(x_1,\ldots,x_n) \suchthat A$ be a query that is applicable to 
$\mathit{DB}$. 
Then a \emph{strongly consistent answer to 
$(x_1,\ldots,x_n) \suchthat A$ with respect to $\mathit{DB}$} is a
$(c_1,\ldots,c_n) \in {\Func{0}(\Sigma)}^n$ such that, 
for each $\Lambda'$ that is $\leq_\Lambda$-minimal in 
$\mathit{Con}(\Xi)$,
\mbox{$\mathit{RT}(R,\Lambda') \scEnt 
       \subst{x_1 \assign c_1}\ldots\subst{x_n \assign c_n}A$} 
is provable.
The elements of $\mathit{Con}(\Xi)$ that are $\leq_\Lambda$-minimal in 
$\mathit{Con}(\Xi)$ are called the \emph{repairs of $\Lambda$}.

The above definition of a strongly consistent answer to a query with 
respect to a database is essentially the same as the definition of a 
consistent answer to a query with respect to a database given 
in~\cite{ABC99a}.
It represents, presumably, the first view on what the repairs of an 
inconsistent database are. 
Other views have been taken in 
e.g.~\cite{LB06a,GM11a,CFK12a,CLP18a,Ber19a}.

\subsubsection*{Decidability}

The following theorem concerns the decidability of being an answer to 
a query.
\begin{theorem}
\label{theorem-decidable-add-add}
Let $R = (\Sigma,\SForm{\Sigma})$ be a relational language, 
let $\mathit{DB} = (R,\Theta,\Xi)$ be a relational database, and
let $(x_1,\ldots,x_n) \suchthat A$ be a query applicable to 
$\mathit{DB}$.
Then it is decidable whether, 
for $(c_1,\ldots,c_n) \in {\Func{0}(\Sigma)}^n$: 
\pagebreak[2]
\begin{itemize}
\item
$(c_1,\ldots,c_n)$ is an answer to 
$(x_1,\ldots,x_n) \suchthat A$ with respect to $\mathit{DB}$;
\item
$(c_1,\ldots,c_n)$ is a consistent answer to 
$(x_1,\ldots,x_n) \suchthat A$ with respect to $\mathit{DB}$;
\item
$(c_1,\ldots,c_n)$ is a strongly consistent answer to 
$(x_1,\ldots,x_n) \suchthat A$ with respect to~$\mathit{DB}$.
\end{itemize}
\end{theorem}
\begin{proof}
Each of these decidability results follows immediately from 
Theorem~\ref{theorem-decidable-add} and the definition of the kind of 
answer concerned.
\qed
\end{proof}
As a corollary of Theorem~\ref{theorem-decidable-add-add}, we have that
the set of answers to a query, the set of consistent answers to a query, 
and the set of strongly consistent answers to a query are computable.

\section{Examples of Query Answering}
\label{EXAMPLES}

For a given database and query applicable to that database, the set of 
all answers, the set of all consistent answers, and the set of all 
strongly consistent answers may be different. 
The examples of query answering given below illustrate this.
The examples are kept extremely simple so that readers that are not 
initiated in the sequent calculus proof system of \foLPif\ can 
understand the remarks made about the provability of sequents.

\subsubsection*{Example~1}

Consider the relational database whose relational language, say $R$, has 
constant symbols $a$ and $b$ and unary predicate symbols $P$ and $Q$, 
whose relational theory is the relational theory of which 
$P(a)$, $P(b)$, and $Q(a)$ are the atomic facts, and whose only 
integrity constraint is $\CForall{x}{\Not (P(x) \CAnd Q(x))}$.
Moreover, consider the query $x \suchthat P(x)$.
Clearly, the set of answers is $\set{a,b}$.

The sets of semi-atomic formulas that are logical consequences of the
relational theory and do not cause the database to be inconsistent are 
$\set{P(b),\Not Q(b)}$ and all its subsets.
We have:
\begin{itemize}
\item
$P(b),\Not Q(b), \mathit{RSA}(R) \scEnt P(a)$ is not provable;
\item
$P(b),\Not Q(b), \mathit{RSA}(R) \scEnt P(b)$ is provable.
\end{itemize}
Hence, the set of consistent answers is $\set{b}$.

The repairs of $\set{P(a),P(b),Q(a)}$ are $\set{P(a),P(b)}$ and
$\set{P(b),Q(a)}$.
We have:
\pagebreak[2]
\begin{itemize}
\item
$\mathit{RT}(R,\set{P(b),Q(a)}) \scEnt P(a)$ is not provable;
\item
$\mathit{RT}(R,\set{P(a),P(b)}) \scEnt P(b)$ is provable;
\item
$\mathit{RT}(R,\set{P(b),Q(a)}) \scEnt P(b)$ is provable.
\end{itemize}
Hence, the set of strongly consistent answers is $\set{b}$.

In this example, the set of all answers differs from the set of all 
consistent answers and the set of all strongly consistent answers, but
the set of all consistent answers and the set of all 
strongly consistent answers are the same.
The repairs of the database are obtained by deletion of atomic facts.

\subsubsection*{Example~2}

Consider the relational database whose relational language, say $R$, has 
constant symbols $a$, $b$, and $c$ and unary predicate symbols $P$ and 
$Q$, whose relational theory is the relational theory of which 
$P(a)$, $P(b)$, $Q(a)$, and $Q(c)$ are the atomic facts, and whose 
only integrity constraint is $\CForall{x}{P(x) \SImpl Q(x)}$.
Moreover, consider the query $x \suchthat P(x)$.
Clearly, the set of answers is $\set{a,b}$.

The sets of semi-atomic formulas that are logical consequences of the
relational theory and do not cause the database to be inconsistent are 
$\set{P(a),\Not P(c),\linebreak[2]Q(a),\Not Q(b),Q(c)}$ and all its subsets.
We have:
\begin{itemize}
\item
$P(a),\Not P(c),Q(a),\Not Q(b),Q(c), \mathit{RSA}(R) \scEnt P(a)$ is 
provable;
\item
$P(a),\Not P(c),Q(a),\Not Q(b),Q(c), \mathit{RSA}(R) \scEnt P(b)$ is 
not provable.
\end{itemize}
Hence, the set of consistent answers is $\set{a}$.

The repairs of $\set{P(a),P(b),Q(a),Q(c)}$ are  
$\set{P(a),P(b),Q(a),Q(b),Q(c)}$ and $\set{P(a),Q(a),Q(c)}$.
We have:
\begin{itemize}
\item
$\mathit{RT}(R,\set{P(a),P(b),Q(a),Q(b),Q(c)}) \scEnt P(a)$ is provable;
\item
$\mathit{RT}(R,\set{P(a),Q(a),Q(c)}) \scEnt P(a)$ is provable;
\item
$\mathit{RT}(R,\set{P(a),Q(a),Q(c)}) \scEnt P(b)$ is not provable.
\end{itemize}
Hence, the set of strongly consistent answers is $\set{a}$.

In this example, like in the previous example, the set of all answers 
differs from the set of all consistent answers and the set of all 
strongly consistent answers, but the set of all consistent answers and 
the set of all strongly consistent answers are the same.
Unlike in the previous example, one of the repairs of the database is 
obtained by deletion of an atomic fact and the other is obtained by 
insertion of an atomic fact.

\subsubsection*{Example~3}

Consider the relational database whose relational language, say $R$, has 
constant symbols $a$, $b$, $c$, $d$, $e$, $f$, and $g$ and unary 
predicate symbol $P$, whose relational theory is the relational theory 
of which
$P(a,b,c)$, $P(a,c,d)$, $P(a,c,e)$, and $P(b,f,g)$ are the atomic facts, 
and whose only integrity constraint is
$\CForall{x,y,z,y',z'}
       {(P(x,y,z) \CAnd P(x,y',z')) \SImpl y = y'}$.
Moreover, consider the query $y \suchthat \CExists{x,z}{P(x,y,z)}$.
Clearly, the set of answers is $\set{b,c,f}$.

The sets of semi-atomic formulas that are logical consequences of the
relational theory and do not cause the database to be inconsistent 
include $\set{P(a,b,c),P(b,f,g)}$ and 
$\set{P(a,c,d),P(a,c,e),P(b,f,g)}$.
We have: 
\begin{itemize}
\item
$P(a,b,c),P(b,f,g), \mathit{RSA}(R) \scEnt \CExists{x,z}{P(x,b,z)}$ is 
provable;
\item
$P(a,c,d),P(a,c,e),P(b,f,g), \mathit{RSA}(R) \scEnt
 \CExists{x,z}{P(x,c,z)}$ 
is provable;
\item
$P(a,b,c),P(b,f,g), \mathit{RSA}(R) \scEnt \CExists{x,z}{P(x,f,z)}$ is 
provable.
\end{itemize}
Because $a$, $d$, $e$, and $g$ are no answers, they cannot be consistent 
answers.
Hence, the set of consistent answers is $\set{b,c,f}$.

The repairs of $\set{P(a,b,c),P(a,c,d),P(a,c,e),P(b,f,g)}$ are 
$\set{P(a,b,c),\linebreak[2] P(b,f,g)}$ and 
$\set{P(a,c,d),P(a,c,e),P(b,f,g)}$.
We have:
\pagebreak[2]
\begin{itemize}
\item
$\mathit{RT}(R,\set{P(a,c,d),P(a,c,e),P(b,f,g)}) \scEnt
 \CExists{x,z}{P(x,b,z)}$ 
is not provable;
\item
$\mathit{RT}(R,\set{P(a,b,c),P(b,f,g)}) \scEnt \CExists{x,z}{P(x,c,z)}$ 
is not provable;
\item
$\mathit{RT}(R,\set{P(a,c,d),P(a,c,e),P(b,f,g)}) \scEnt
 \CExists{x,z}{P(x,f,z)}$ 
is provable;
\item
$\mathit{RT}(R,\set{P(a,b,c),P(b,f,g)}) \scEnt \CExists{x,z}{P(x,f,z)}$ 
is provable.
\end{itemize}
Because $a$, $d$, $e$, and $g$ are no answers, they cannot be 
strongly consistent answers.
Hence, the set of strongly consistent answers is $\set{f}$.

In this example, unlike in the previous two examples, the set of all 
answers and the set of all consistent answers are the same, but the set 
of all consistent answers differs from the set of all 
strongly consistent answers.
Like in the first example, the repairs of this database are obtained by 
deletion of atomic facts.

\section{Some remarks about consistent query answering}
\label{REMARKS}

The definition of a consistent answer to a query with respect to a 
database given in Section~\ref{QUERY-ANSWERING} simply accepts that a 
database is inconsistent and excludes the source or sources of 
inconsistency from being used in consistent query answering.
Several considerations underlying this definition are mentioned in the
next two paragraphs.

Seeing the extensional nature of the atomic facts of a database and the
intensional nature of its integrity constraints, it is natural to 
consider the presence or absence of atomic facts in a database that 
causes inconsistency with its integrity constraints suspect and 
consequently not to use it in answering a query with respect to the 
database.
The plain choice not to use the source or sources of inconsistency in 
answering a query does not result in additional choices to be made.

The only accepted alternative to deal with an inconsistent database is 
to base the answers on consistent databases, called repairs, obtained by 
deletion and/or addition and/or alteration of atomic facts from the 
inconsistent database that differ to a minimal extent from the 
inconsistent database.
This alternative requires rather artificial choices to be made 
concerning, among other things, the kinds of changes (deletions, 
additions, alterations) that may be made to the original database and
what is taken as the extent to which two databases differ.

The definition of a consistent answer to a query with respect to a 
database given in Section~\ref{QUERY-ANSWERING} is reminiscent of the 
definition of a consistent answer to a query with respect to a database 
given in~\cite{Bry97a}.
That paper is, to my knowledge, the first paper in which consistent 
query answering in inconsistent databases is considered. 
The definition of consistent query answer given in that paper is 
based on provability in a natural deduction proof system of first-order
minimal logic, a paraconsistent logic that is much less close to 
classical logic than \foLPif.

What is missing in~\cite{Bry97a} is a semantics with respect to which 
the presented proof system is sound and complete. 
This leaves it somewhat unclear how the logical versions of the 
relevant notions (relational database, query, etc.) defined in that 
paper are related to their standard version.
The Kripke semantics of the propositional fragment of minimal logic that
can  be found in various publications leaves this unclear as well.

The definition of a strongly consistent answer to a query with 
respect to a database given in this section is essentially the same as 
the definition of a consistent answer to a query with respect to a 
database given in~\cite{ABC99a}.
It is, to my knowledge, the first definition of a consistent answer 
based on the idea that an answer is consistent if it is an answer with 
respect to any minimally repaired version of the original database.
Different views of what is a minimally repaired version of a database
are plausible.
Views that differ from the original one have been considered in 
e.g.~\cite{LB06a,GM11a,CFK12a,CLP18a,Ber19a}.

In~\cite{Bry97a}, the definition of a consistent answer is based on the 
idea that a (usually large) part of an inconsistent database is 
consistent and that a consistent answer is simply an answer with respect 
to the consistent part of the database. 
From the viewpoint taken in~\cite{ABC99a}, this means that only one 
repair is considered.
Because there is in general more than one repair of a database, this is 
called a shortcoming in~\cite{Cho04a}. 
However, the implicit assumption that it is necessary to use the 
auxiliary notion of a repair in defining the notion of a consistent 
answer is nowhere substantiated.

\section{Concluding Remarks}
\label{CONCLUSIONS}

This paper builds heavily on the following views related to relational 
databases and consistent query answering:
\begin{itemize}
\item
the proof-theoretic view of \cite{Rei84a} on what is a relational 
databases, a query applicable to a relational database, and an answer to 
a query with respect to a consistent relational database;
\item
the view of \cite{Bry97a} on what is a consistent answer to a query with 
respect to an inconsistent relational database;
\item
the view of \cite{ABC99a} on what is a consistent answer to a query with 
respect to an inconsistent relational database.
\end{itemize}
The view of Reiter~\cite{Rei84a} has been combined with the view of 
Bry~\cite{Bry97a} as well as with the view of Arenas et al~\cite{ABC99a} 
and adapted to the setting of the paraconsistent logic \foLPif.
This has led to one coherent view on relational databases and consistent 
query answering expressed in a setting that is more suitable to this end
than classical logic or minimal logic.

The notion of a relational theory can be generalized by allowing its 
basis to be a set of Horn clauses and adapting the completion axioms as
sketched in~\cite{GMN84a}.
This generalization gives rise to a generalization of the notion of a
relational database that is generally known as the notion of a definite 
deductive database. 
The definitions of an answer, a consistent answer, and a strongly 
consistent answer given in this paper are also applicable to this 
generalization of the notion of a relational database.
Further generalizion of the notion of an indefinite deductive database 
is a different matter.

The presented sequent calculus proof system of \foLPif, which is sound 
and complete with respect of the given three-valued semantics of 
\foLPif, is new.

\bibliographystyle{splncs03}
\bibliography{PCL}

\end{document}